\documentclass[aps,prl,twocolumn,showpacs,superscriptaddress,groupedaddress]{revtex4}
\usepackage{graphicx}  
\usepackage{dcolumn}   
\usepackage{bm}        
\usepackage{amssymb}   
\usepackage{amsmath}   
\usepackage{subfigure}

\def\QE{{\sc Quantum ESPRESSO}}

\begin{document}
\widetext
\title{A Local Representation of the Electronic Dielectric Response Function}

\author{Xiaochuan Ge}
\author{Deyu Lu} \email{dlu@bnl.gov}
\affiliation{Center for Functional Nanomaterials, Brookhaven National Laboratory
, Upton, New York 11973, United States}
\date{\today}

\begin{abstract}
We present a local representation of the electronic dielectric
response function, based on a spatial partition of the dielectric
response into contributions from each Wannier function using a
generalized density functional perturbation theory. This procedure is
fully \emph{ab initio}, and therefore allows us to rigorously define
local metrics, such as ``bond polarizability'', on Wannier centers.
We show that the locality of the response function is determined by
the locality of three quantities: Wannier functions of the occupied
manifold, the density matrix, and the Hamiltonian matrix. In systems
with a gap, the bare dielectric response is exponentially localized,
which supports the physical picture of the dielectric response
function as a collection of interacting local response that can be
captured by a tight-binding model.
\end{abstract}

\maketitle
The screened electronic dielectric response function (EDRF), $\chi$,
is a fundamental physical quantity that captures many-electron
correlation effects.  From a microscopic point of view, $\chi$ relates
the perturbation from an external potential at ${\bf r'}$ to the
electronic density response at ${\bf r}$. This intrinsic non-local
character has precluded a compact local representation of $\chi$ in
electronic structure theory; it has been traditionally represented by
huge matrices in either real or reciprocal space~\cite{ONID2002}. This
cumbersome matrix representation of $\chi$ has become a major
computational bottleneck to accurately predict electron correlation
energy and electronic excitation spectra. More importantly, the
physical interpretation of $\chi$ is largely limited to its
macroscopic average. A robust, microscopic theory that describes the
local characteristics (e.g., shape, strength and decay rate) is needed
to unravel the underlying physical nature of EDRFs.

Empirical methods have been used to partition EDRFs to obtain
polarizabilities or effective van der Waals $C_6$ dispersion
coefficients of atoms inside either a molecule or a
solid~\cite{Wu2002, Tkatchenko2009}. On the other hand, non-empirical
methods often rely on extra approximations to partition, e.g., the
molecular polarizability, into so-called ``distributed
polarizabilities''~\cite{STON1985}. Such procedures typically involve
partitioning the volume~\cite{ANGY1994} or the basis
space~\cite{LESU1993} of a molecule, or fitting the point-to-point
polarizabilities computed on a grid around a molecule~\cite{WILL2003},
with known drawbacks including large charge-flow terms that are hard
to localize, strong basis set dependence, and high computational
cost~\cite{MISQ2006}.

The Wannier function (WF) representation~\cite{WANN1937} is a natural
choice to describe chemical bonds using either ``Boys orbitals'' for
molecules~\cite{BOYS1960} or the maximally localized Wannier functions
(MLWFs) for the solid-state
equivalent~\cite{Marzari1997,SOUZ2001,Marzari2012}. However, the link
between local EDRFs and WFs is obscured by the fact that EDRFs concern
electron-hole pairs rather than electronic orbitals
alone. Silvestrelli~\cite{SILV2008} proposed to define $C_6$
coefficients on Wannier centers using empirical models, assuming that
WFs have the $s$-symmetry. Giustino and Pasquarello~\cite{GIUS2005}
introduced the local dielectric permittivity in layered systems, based
on local dipole moments derived from each Wannier function using the
Berry-phase theory of the polarization~\cite{REST1994}. However, this
approach requires separate calculations under finite external fields,
unsuitable to study the spatial decay rate and the dynamic
response. Lu \emph{et al.}~\cite{LU2008} applied a simultaneous
diagonalization algorithm to directly localize the eigenvectors of
EDRFs. Despite the observed trend in the locality~\cite{LU2008}, the
chemical nature of localized EDRFs was not determined precisely.

In this Letter, we propose a local representation for microscopic
EDRFs using a generalized density functional perturbation theory
(DFPT)~\cite{Baroni2001}. While the conventional theory is formulated
on the eigenstates of the Kohn-Sham (KS) Hamiltonian, we generalize
the DFPT to any orthogonal basis set that spans the occupied state
manifold. A convenient choice adopted in this work is the
MLWF~\cite{Marzari1997, SOUZ2001}, as it can provide insightful
interpretations regarding chemical bonds.  Because this method is fully
\emph{ab initio}, it ensures accuracy and transferability.  

First we generalize DFPT for the bare EDRF, $\chi^0$, which is the
building block in linear response theory.  Under a perturbation in the
self-consistent potential, $e^{-i \omega t}\Delta V_{s}({\bf r})$, the
response density can be calculated through $\chi^0$ as
\begin{equation}
  \Delta \rho (\omega; {\bf r}) = \int d{\bf r'} \; \chi^0
  (\omega;{\bf r},{\bf r'})\, \Delta V_s({\bf r'}).
  \label{unscreened-DRF}
\end{equation}
In the following, we adopt the shorthand notation: $\Delta \rho =
\chi^0\, \Delta V_s$. $\chi$ can be solved from $\chi^0$ through
Dyson's equation, $\chi = \chi^0 + \chi^0 \, K\, \chi$, where $K=v_c +
K_{xc}$ with $v_c$ and $K_{xc}$ being Coulomb and exchange-correlation
kernels, respectively~\cite{FETT2003}. For periodic systems,
Eq.~\ref{unscreened-DRF} is often solved for individual Fourier
components with wave vectors ${\bf q}$, $\Delta V^{\bf q}_s({\bf r})=e^{i
{\bf q \cdot r}}\Delta v_s({\bf r})$. The response density matrix is
given by
\begin{equation}
  \Delta \rho_{\bf q} =
    \frac{2}{\Omega} \sum_{v{\bf k}} 
    \left |  \Delta \psi^{{\bf k}+\bf {q}}_{v\pm} \right \rangle 
    \Bigl \langle \psi_v^{\bf k} \Bigr |,
  \label{res-den}
\end{equation}
where the variation of the KS orbital, $|\Delta \psi^{{\bf k}+\bf
{q}}_{v\pm}\rangle$, is the solution of the Sternheimer
equation~\cite{Baroni2001},
\begin{equation}
  (\varepsilon_v^{\bf k}-H-\alpha P_v^{{\bf k}+\bf {q}} \pm \omega)
  \left | \Delta \psi_{v\pm}^{{\bf k}+\bf {q}} \right \rangle =
  P_c^{{\bf k}+\bf {q}} \Delta V^{\bf q}_s \left | \psi_v^{\bf k} \right
  \rangle.
  \label{stern-KS}
\end{equation}
Here $\alpha P_v^{{\bf k}+\bf {q}}$ makes Eq.~\ref{stern-KS}
non singular; $P_v^{\bf k+q}$ and $P_c^{\bf k+q}$ are projectors onto
the occupied and unoccupied state manifolds at momentum $\bf{k+q}$,
which are introduced to avoid the explicit reference to the unoccupied
states~\cite{Baroni2001}.

It is trivial to partition $\Delta \rho_{\bf q}$ in Eq.~\ref{res-den}
in the energy domain into contributions from individual KS orbitals at
given \{$v$, ${\bf k}$\} and {\bf q}, and the corresponding linear
equations in Eq.~\ref{stern-KS} are decoupled. Alternatively, a real
space partition of EDRFs can be achieved through a generalized DFPT in
the WF representation. Following the notations in
Ref.~\cite{Marzari1997}, we define WFs and their first order
perturbations as
\begin{equation}
  \begin{aligned} W_{{\bf R}n}({\bf r}) &=
  \frac{\Omega}{(2\pi)^3}\int_{BZ} d{\bf k} e^{-i{\bf k}\cdot {\bf R}}
  \sum_{m=1}^J U^{({\bf k})}_{mn} \psi_{m}^{\bf k} ({\bf r}), \\
  \Delta W^\pm_{{\bf R}n}({\bf r}) &=
  \frac{\Omega}{(2\pi)^3}\int_{BZ} d{\bf k} e^{-i{\bf
  k}\cdot {\bf R}} \sum_{\bf q} \sum_{m=1}^J U^{({\bf k})}_{mn} \Delta
  \psi_{m\pm}^{\bf k+q} ({\bf r}),
  \label{WF-RWF}
  \end{aligned} 
\end{equation}
where $J$ is the number of orbitals used to construct WFs, and unitary
matrices $U^{({\bf k})}$ minimize the spatial spreads of the WFs
labeled by lattice vector ${\bf R}$ and band index
$n$~\cite{Marzari1997}. For simplicity, we will focus on the the
static limit and drop the superscripts $+$ and $-$; extension to the
dynamic case is straightforward. Because $\rho$ and, thus, $\Delta
\rho$ are invariant under the unitary transformation of occupied
states, Eq.~\ref{res-den} can be rewritten in the Wannier
representation as
\begin{equation}
        \Delta \rho = \frac{4}{\Omega}\sum_{{\bf R}n} \left | \Delta
          W_{{\bf R}n} \right \rangle \left \langle W_{{\bf R}n}
          \right | .
  \label{res-den-WF}
\end{equation}

Applying $U^{({\bf k})}$ to both sides of Eq.~\ref{stern-KS} and
integrating over ${\bf k}$, one obtains the generalized Sternheimer
equation in the Wannier representation as
\begin{equation}
  \sum_{{\bf R}'n'}(\tilde \varepsilon_{{\bf R}n,{\bf R}'n'}-H-\alpha
   P_{v}) \left | \Delta W_{{\bf R}'n'} \right \rangle = P_{c} \Delta
   V_s \left | W_{{\bf R}n} \right \rangle .
  \label{stern-WF}
\end{equation}
Because WFs are not eigenstates of the KS Hamiltonian, unlike
Eq.~\ref{stern-KS}, Eq.~\ref{stern-WF} yields a set of coupled
equations due to the hopping integral terms, $\tilde \varepsilon_{{\bf
R}n,{\bf R}'n'}= \langle W_{{\bf R}n} | H | W_{{\bf R}'n'} \rangle$
(${\bf R}n \ne {\bf R}'n'$). It follows that
\begin{equation}
      \begin{aligned}
        \left | \Delta W_{{\bf R}n} \right \rangle &= \sum_{{\bf
        R}'n'} \left | \Delta W_{ {\bf R}n, {\bf R}'n'} \right \rangle
        \\ &\equiv \sum_{{\bf R}'n'} [\tilde \varepsilon -(H+\alpha
        P_{v})\,I]^{-1}_{{\bf R}n,{\bf R}'n'} P_c \Delta V_s \left |
        W_{{\bf R}'n'}\right \rangle ,
      \end{aligned}
  \label{stern-WF-sol}
\end{equation}
where $I$ is an $N_W \times N_W$ identity matrix with $N_W$ being the
number of occupied WFs. $\Delta W_{{\bf R}n,{\bf R}'n'}$ denotes the
variation of $W_{{\bf R}n}$ caused by the perturbation at $W_{{\bf
R}'n'}$. Combining Eqs.~\ref{res-den-WF} and~\ref{stern-WF-sol}, one
can expand $\chi^0$ in terms of two-body partial response functions,
$\chi^0 = \sum_{{\bf R}n,{\bf R}'n'} \chi^0_{{\bf R}n,{\bf R}'n'}$,
with corresponding density response given by $\Delta \rho_{_{{\bf
R}n,{\bf R}'n'}} \equiv \frac{4}{\Omega} | \Delta W_{{\bf R}n,{\bf
R}'n'} \rangle \langle W_{{\bf R}n} |$. A formal real space partition
can be established by contracting $\chi^0_{{\bf R}n,{\bf R}'n'}$ into
one-body variables,
\begin{equation}
  \begin{aligned}
  \chi^0 ({\bf r,r'})&=\sum_{{\bf R}n} \chi^0_{{\bf R}n} ({\bf r,r'}), \\
  \chi^0_{{\bf R}n}({\bf r,r'}) &= \frac{1}{2}\sum_{{\bf R}'n'} \left [\chi^0_{{\bf
  R}n,{\bf R}'n'}({\bf r,r'}) + \chi^0_{{\bf R}'n',{\bf R}n}({\bf r,r'}) \right ]. \label{Erpart}
  \end{aligned}
\end{equation}
$\chi^0_{{\bf R}n}$ is Hermitian by construction. It contains not only
on-site terms, but also the charge-flow (${\bf R}'n'\ne {\bf R}n$)
into and out of the local orbital, whose magnitude determines the
extent of the locality of $\chi^0$. Eq.~\ref{Erpart} is the central
equation of our theory, which together with the locality analysis
below formally establishes the local representation of EDRFs. The real
space partition can be achieved similarly for $\chi$ through partial
response functions, $\chi_{{\bf R}n} = \frac{1}{2}\sum_{{\bf R}'n'}
\left (\chi_{{\bf R}n,{\bf R}'n'} + \chi_{{\bf R}'n',{\bf R}n}\right
)$, where
\begin{eqnarray}
\chi_{{\bf R}n,{\bf R}'n'} &=& \sum_{{\bf R}''n''}\sum_{m=0}^{\infty}
\chi^0_{{\bf R}n,R''n''} \left( K \,\chi^0\right)^m_{{\bf
R''}n'',{\bf R}'n'}, \nonumber \\ 
\chi_{{\bf R}'n',{\bf R}n} &=&
\sum_{{\bf R}''n''}\sum_{m=0}^{\infty} \left ( \chi^0\,
K\,\right )^m_{{\bf R}'n',R''n''}\, \chi^0_{{\bf R''}n'',{\bf R}n},\label{Erpartchi}
\end{eqnarray}
are solved self-consistently with Eq.~\ref{stern-WF}.  We emphasize
that there is a critical distinction between
Eqs.~\ref{Erpart}-\ref{Erpartchi} and existing approximate partition
methods. In our method, once a Wannier localization procedure is
chosen, the subsequent real space EDRF partition is exact.

To demonstrate the concept of the local EDRF, we partition static
molecular polarizabilities of acetylene ($C_2H_2$), ethylene
($C_2H_4$), and ethane ($C_2H_6$) into bond polarizabilities at
Wannier centers. The major difference between these molecules is the
$C$-$C$ bond order: single bond in $C_2H_6$, double bond in $C_2H_4$,
and triple bond in $C_2H_2$.  We compare the standard mean screened
polarizability ($\bar \alpha=\frac{1}{3}tr(\hat{\bf r} \chi \hat{\bf
r}')$) with the unscreened one ($\bar{\alpha}^0=\frac{1}{3}tr(\hat{\bf
r} \chi^0 \hat{\bf r}'$)) to gain insight into the local field
effect. All the calculations were performed at the random phase
approximation level using Wannier90~\cite{wannier90} and a modified
version of \QE~\cite{Giannozzi2009}; computational details are given
in the Supplemental Material (SM)~\cite{SupMat}.
\begin{table}[ht]
\caption{\label{tab-c2hn} Mean static molecular and bond
polarizabilities (in Bohr$^3$) of $C_2H_6$, $C_2H_4$, and
$C_2H_2$. Multipliers in bond polarizabilities indicate the
degeneracy; unscreened bond polarizabilities are shown in the
parentheses.}
\begin{ruledtabular}
\begin{tabular}{c|ccc|cc}
&Total&C-H&C-C&$\sigma_{CC}$&$\pi_{CC}$\\
\hline
$C_2H_2$ &24.65 &2$\times$3.11(4.36) &18.43 & 0.60(1.55) & 2$\times$8.92(16.26) \\
$C_2H_4$ &29.41 &4$\times$3.91(5.73) & 13.76 & 1.73(3.30) & 12.02(21.28)   \\
$C_2H_6$ &31.35 &6$\times$4.54(6.60) &4.09  & 4.09(6.25) &   \\
\end{tabular}
\end{ruledtabular}
\end{table}

Within $C_2H_2$, as shown in Table~\ref{tab-c2hn},
$\bar\alpha^0(\pi_{CC})>\bar\alpha^0(CH)>\bar\alpha^0(\sigma_{CC})$. This
behavior is a direct outcome of the electronic structure, because
$\pi_{CC}$ is closest to the lowest unoccupied molecular orbital
(LUMO) in energy (i.e., most reactive), and $\sigma_{CC}$ is farthest
from LUMO (i.e., least reactive) as shown by the projected density of
states in Fig. S2 in SM. The same trend holds for all three
molecules. Among different molecules, unscreened bond polarizabilities
increase in the order of $C_2H_2$, $C_2H_4$, and $C_2H_6$, partially
due to the reduced bonding-antibonding splitting with increased bond
lengths ($d_{CH}$: 1.000, 1.018, 1.024; $d_{CC}$: 1.000, 1.102, 1.267,
both normalized by the bond lengths of $C_2H_2$). Another dominating
factor in the molecular polarizability is the number of bonds, i.e.,
the degeneracy.  On the other hand, $\bar\alpha$ is always smaller
than $\bar\alpha^0$, because of the screening effect.  We define
$\epsilon_{eff}=\bar\alpha^0 / \bar\alpha$ as a measure of the local
screening strength that includes both intra- and inter-bond screening
effects. $\epsilon_{eff}$ is highly heterogeneous in these molecules,
and the largest values arise from the strongly overlapping
$\sigma_{CC}$ and $\pi_{CC}$ bonds. Consequently,
$\epsilon_{eff}(\sigma_{CC})$ (1.9 and 2.6 ) and
$\epsilon_{eff}(\pi_{CC})$ (1.8) in $C_2H_2$ and $C_2H_4$ are
significantly larger than $\epsilon_{eff}(\sigma_{CH})$ ($1.4 \sim
1.5$) and $\epsilon_{eff}(\sigma_{CC})$ in $C_2H_6$ (1.5).

\begin{figure}[b]
\includegraphics[width=3.0 in]{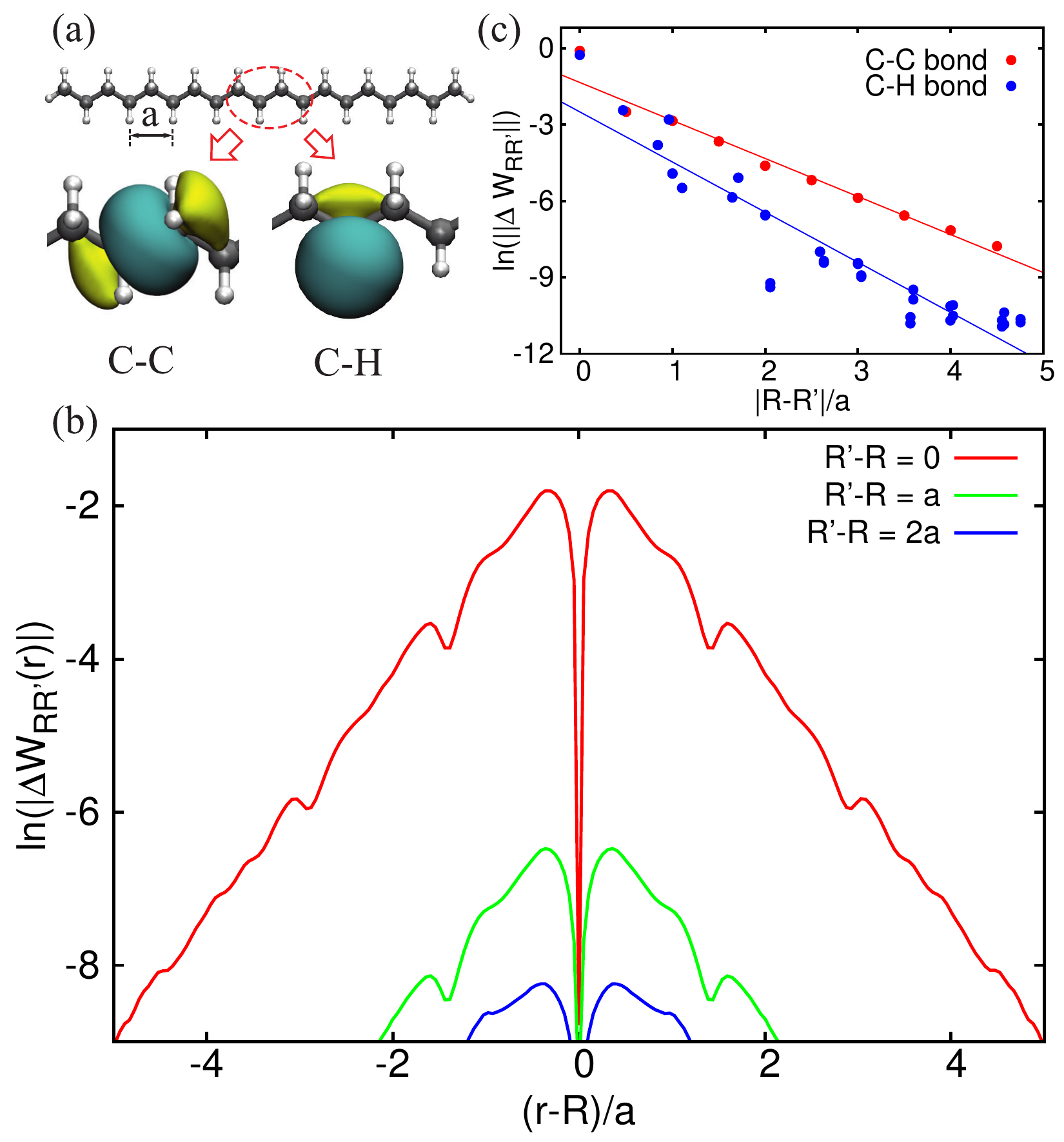}
\caption{ \label{fig-locality} Locality of $\Delta W_{{\bf RR}'}({\bf
r})$ in an ethylene oligomer ($C_{19}H_{40}$). (a) Wannier orbitals of
C-C and C-H bonds. (b) The exponential decay of $\Delta W_{{\bf
RR}'}({\bf r})$ as a function of $|{\bf r}-{\bf R'}|$ (only shown for
C-H bond) and (c) of $||\Delta W_{{\bf RR}'}||$ as a function of
$|{\bf R-R'}|$. Solid lines in (c) indicate a linear fit.}
\end{figure}

For systems with a finite gap, it has been proved that (a) $W_{{\bf
R}n}({\bf r})$ decays exponentially with $|{\bf r}-{\bf
R}|$~\cite{Kohn1959,PANA2013}; (b) the Hamiltonian matrix $\tilde
\varepsilon_{{\bf R}n,{\bf R}'n'}$ decays exponentially with $|{\bf
R}-{\bf R}'|$~\cite{He2001}; and (c) the density matrix $\langle {\bf
r} | P_v | {\bf r}' \rangle $ decays exponentially with $|{\bf r}-{\bf
r}'|$~\cite{Goedecker1998, Ismail-Beigi1999, He2001}. The locality of
$\Delta W_{{\bf R}n,{\bf R}'n'} ({\bf r})$ can be derived accordingly.
(i) For a regular potential that is not exponentially divergent, $P_c
\Delta V_{s} W_{{\bf R'}n'}({\bf r})$ decays exponentially with $|{\bf
r}-{\bf R'}|$.  This can be easily understood from the locality of
$W_{{\bf R}n}({\bf r})$ and $ P_v $, as $P_c = I - P_v$.  (ii)
$[\tilde \varepsilon -(H + \alpha P_v)I]^{-1}_{\bf RR'}$ decays
exponentially with $ |{\bf R-R'}|$, a direct consequence of the
locality of $\tilde \varepsilon_{{\bf R}n,{\bf R}'n'}$ \footnote{
Since its off-diagonal elements decay exponentially with distance,
$\tilde\varepsilon$ can be truncated beyond the distance that is
several times of the decay length, so that $(H + \alpha P_v)I -\tilde
\varepsilon$ becomes a positive definite block banded matrix. It has
been proved that the inverse of such a matrix has exponentially
decayed off-diagonal elements~\cite{Demko1984}.}.  (iii) Putting (i)
and (ii) together, we conclude that $\Delta W_{{\bf R}n,{\bf R}'n'}
({\bf r})$ decays exponentially with $|{\bf r}-{\bf R'}|$ for a given
${\bf R}$ and ${\bf R'}$ pair, and its two-norm, $||\Delta W_{{\bf
R}n,{\bf R}'n'}||= \sqrt{\int d{\bf r} |\Delta W_{{\bf R}n,{\bf R}'n'}
({\bf r})|^2}$, decays exponentially with $|{\bf R-R'}|$.

To validate these arguments, we considered an ethylene oligomer
($C_{19}H_{40}$ with unit length $a$) containing two types of WFs, C-C
and C-H bonds, as shown in Fig.~\ref{fig-locality}a. To quantify the
locality of $\Delta W_{{\bf R}n{\bf R}'n'}$, we consider a linear
potential $V_s ({\bf r}) = x$ applied along the molecule. Both $\Delta
W_{{\bf R}n{\bf R}'n'} ({\bf r})$ (Fig.~\ref{fig-locality}b) and
$||\Delta W_{{\bf R}n,{\bf R}'n'}||$ (Fig.~\ref{fig-locality}c)
clearly exhibit an exponential decay. While the decay length of the
former ($0.6\sim1.0a$) is governed by the locality of the WFs
($0.3\sim 0.5a$) and the density matrix, the latter ($0.5 \sim 0.7a$)
is controlled by the locality of $\tilde \varepsilon_{{\bf R}n{\bf
R'}n'}$ (see Fig.~S4 in SM).

\begin{figure}[t]
  \includegraphics[width=3.0 in]{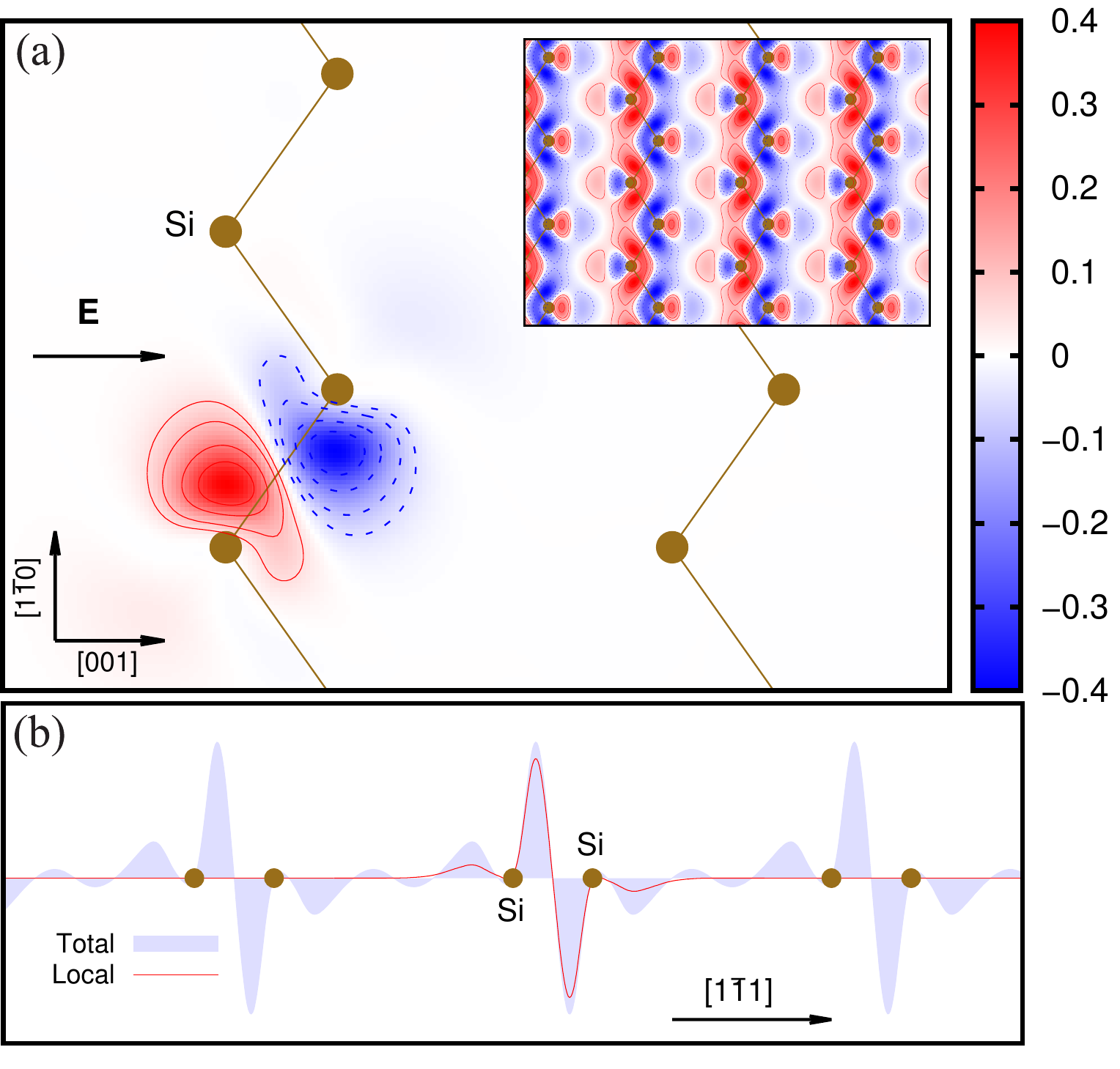}
  \caption{ \label{fig-si-response} The (local) density response of
  silicon under a uniform electronic field in the [001] direction. (a)
  Contour plot of the local density response in the [110] plane of a
  Si-Si bond (inset: total density response). (b) Density response
  profile along the Si-Si bond.}
\end{figure}

The strong spatial localization of the response density also exists in
3D crystals, as shown in bulk silicon in Fig.~\ref{fig-si-response}.
Under a uniform electronic field, although the total response density
is delocalized, $82$\% local density response along the
[$1\bar{1}1$] direction is confined within one Si-Si bond (see
Fig.~\ref{fig-si-response}b).

An important outcome of the exponential localization of $W_{{\bf
R}n}({\bf r})$ and $\Delta W_{{\bf R}n,{\bf R}'n'({\bf r})}$ is that
$\chi^0_{{\bf R}n,{\bf R}'n'}$ is exponentially localized. It supports
the physical picture of $\chi^0$ as a linear combination of coupled
local response modes centered on WFs. Consequently, efficient
algorithms can be developed to compute $\chi^0$, taking advantage of
its sparse representation in the real space. To prove this point, we
construct a tight-binding model of $\chi^0$ for silicon with a local
basis set, and compute its eigenvalue spectrum in the Brillouin zone
using the dielectric band structure (DBS) interpolation, in analogy to
the electronic band structure interpolation using Wannier
functions~\cite{SOUZ2001}. To our best knowledge, this is the first
demonstration of DBS interpolation for a covalent bonded crystal.

We first construct local basis set of $\chi^0$ denoted by $\left |
\xi^{{\bf R}n}_i\right \rangle$, as approximate eigenvectors of
$\chi^0_{{\bf R}n{\bf R}n}$, where hopping terms
$\tilde{\epsilon}_{{\bf R}n,{\bf R}'n'}$ (${\bf R}n\ne {\bf R}'n'$)
are switched off. This procedure decouples equations in
Eq.~\ref{stern-WF}, making them easy to solve computationally. Then we
calculate the hopping matrix $X$ and overlap matrix $S$ in real space,
$X_{ij}^{nn'}({\bf R}) = \left \langle \xi_i^{0n} | \chi^0 |
\xi_j^{{\bf R}n'} \right \rangle$, and $S_{ij}^{nn'}({\bf R}) = \left
\langle \xi_i^{0n} | \xi_j^{{\bf R}n'} \right \rangle$.  Finally we
Fourier transform them into momentum space,
\begin{eqnarray}
    \tilde X({\bf q}) &=& \sum_{\bf R} X({\bf R}) e^{i{\bf q}\cdot{\bf R}}, \nonumber \\
    \tilde S({\bf q}) &=& \sum_{\bf R} S({\bf R}) e^{i{\bf q}\cdot{\bf R}}.
\end{eqnarray}
The interpolated DBS of $\chi^0$ is the solution of the generalized
eigenvalue problem in non-orthogonal basis:
 \begin{equation}
   \tilde X({\bf q}) {\vec v} = \lambda ({\bf q}) \tilde S({\bf q}) {\vec v}.
 \end{equation}

\begin{figure}[t]
\includegraphics[width=3 in]{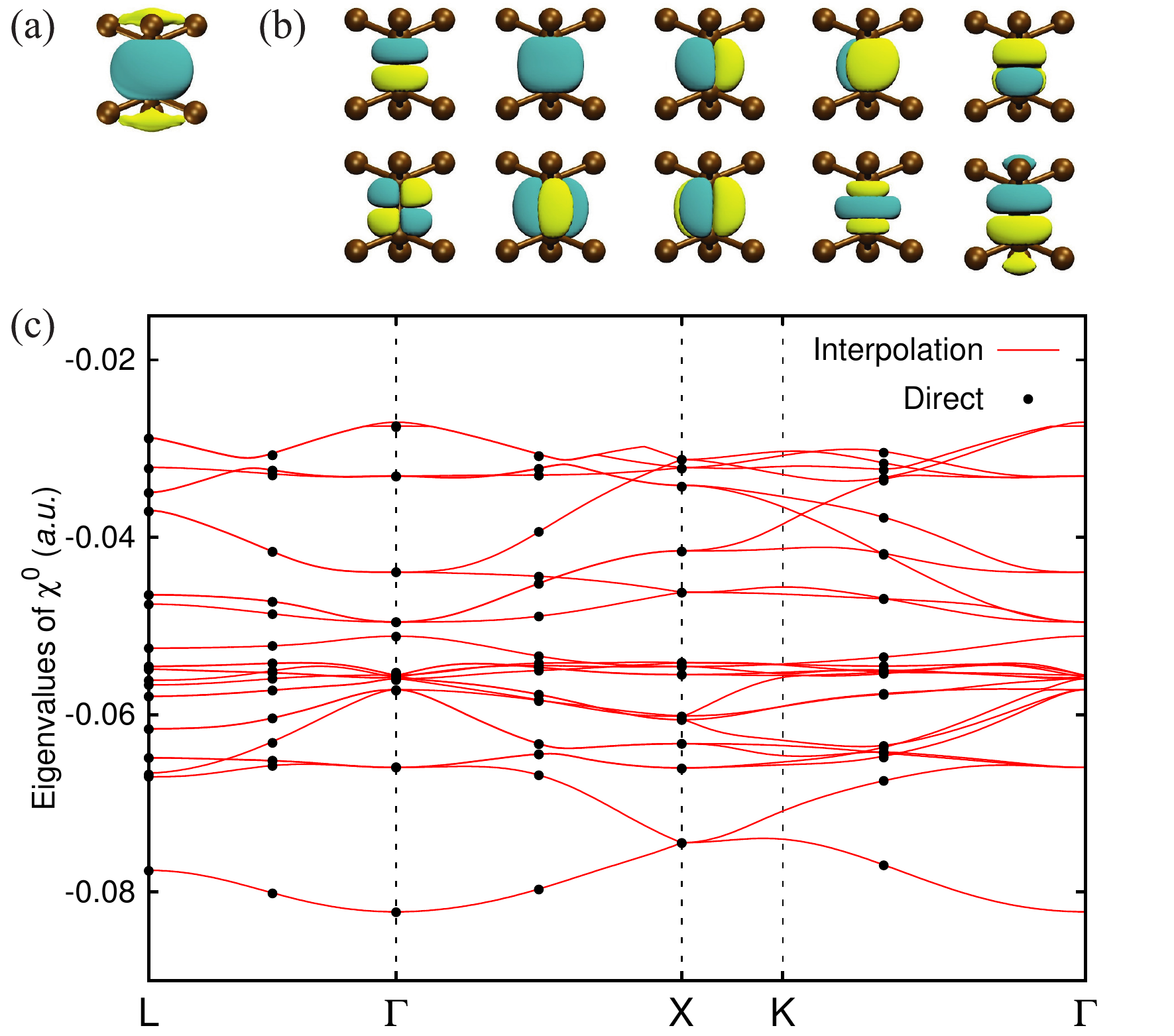}
\caption{\label{fig-dbs} DBS ($\chi^0$) interpolation for bulk silicon
using local basis.  (a) The WF in bulk silicon.  (b) The first 10
eigenmodes of the local response function.  (c) Comparison of the DBS
obtained from a direct calculation with interpolation using the lowest
25 basis functions per WF.}
\end{figure}

We demonstrated the DBS interpolation method with bulk silicon using a
$4\times4\times4$ $k$-mesh for the direct calculation, and a
$4\times4\times4$ super cell with $\Gamma$-point sampling to construct
the local basis set, $X$ and $S$ in real space~\footnote{We have
verified that a $k$-point implementation of DBS interpolation with a
$4\times4\times4$ $k$-mesh would yield the same results, although the
$\Gamma$-point implementation is more straightforward.}. Twenty five
local eigenmodes per WF are sufficient to reach the numerical convergence
of the first 25 bands of the total $\chi^0$.
The Wannier orbitals and the first ten dielectric basis functions of
bulk silicon are shown in Fig.~\ref{fig-dbs}a and
\ref{fig-dbs}b. Excellent agreement (maximum absolute error:
$2.6\times10^{-4} a.u.$) was found between the direct calculation and
interpolated DBS in Fig. \ref{fig-dbs}c, which highlights the validity
and accuracy of the tight-binding model of $\chi^0$.

The local basis set is the key to the efficient construction of
$\chi^0$. Thanks to the exponential localization of $\chi^0_{{\bf
R}n,{\bf R}'n'}$, Eq.~\ref{stern-WF} in principle can be solved within
a subspace containing $n_{n}$ neighboring Wannier orbitals of the
local perturbation. Since $n_{n}$ is system size independent, the
computational cost grows as $O(N^2ln(N))$ with $N$ being the size of
the system, which is a tremendous improvement over the standard
$O(N^4)$ scaling (see SM). This quadratic scaling method is a
promising starting point to develop low scaling algorithms for excited
state problems, where the evaluation of $\chi^0$ is the major
bottleneck to construct, e.g., $\epsilon_{RPA}=1-v\,\chi^0$ or
$\chi$~\cite{LU2009, UMAR2010,ROCC2010}.

The theoretical framework of local EDRFs can be extended to metallic
systems, as DFPT can treat metallic systems in
general~\cite{Baroni2001}, but the decay rate of key quantities can
behave qualitatively differently from semiconductors or insulators.
In 1D free electrons, the Wannier orbital of the occupied portion of
bands decays at $r^{-1}$~\cite{He2001}. For Wannier orbitals of
disentangled bands, e.g., narrow transition metal $d$-bands, both
numerical evidence and the analogy with the isolated composite case
suggest the possibility of the exponential
localization~\cite{Marzari2012}. On the other hand, the decay of
density matrix is expected to be algebraic at zero temperature, and
exponential at finite
temperature~\cite{Ismail-Beigi1999,Goedecker1998}.  The locality of
EDRFs in metallic systems is therefore more subtle, and warrants
further investigation.

In conclusion, we present a local representation of EDRFs based on the
concept of the dielectric response of WFs. This method allows us to
perform fully \emph{ab initio} real space partition of EDRFs, and
analyze excited state properties, e.g., polarizability, in terms of
chemical bonds. In systems with a gap, we proved that the bare
response function, $\chi^0_{{\bf R}n,{\bf R}'n'}$, decays
exponentially in real space. This ``near-sightness'' is central to the
physical understanding of EDRFs and the development of low scaling
algorithms for excited state problems.

This work was performed at the Center for Functional Nanomaterials,
which is a U.S. DOE Office of Science User Facility, at Brookhaven
National Laboratory under Contract No. DE-SC0012704.

\bibliographystyle{h-physrev}

\end{document}


\title{A Local Representation of the Electronic Dielectric Response Function\\Supplemental Material}

\author{Xiaochuan Ge}
\author{Deyu Lu}
\affiliation{Center for Functional Nanomaterials, Brookhaven National Laboratory
, Upton, New York 11973, United States}
\date{\today}
\maketitle

\section{Electronic Structure of $C_2H_n$, ($n=2, 4, 6$)}
\begin{figure}[ht]
\begin{minipage}[t]{0.22\textwidth}
\includegraphics[scale=0.28]{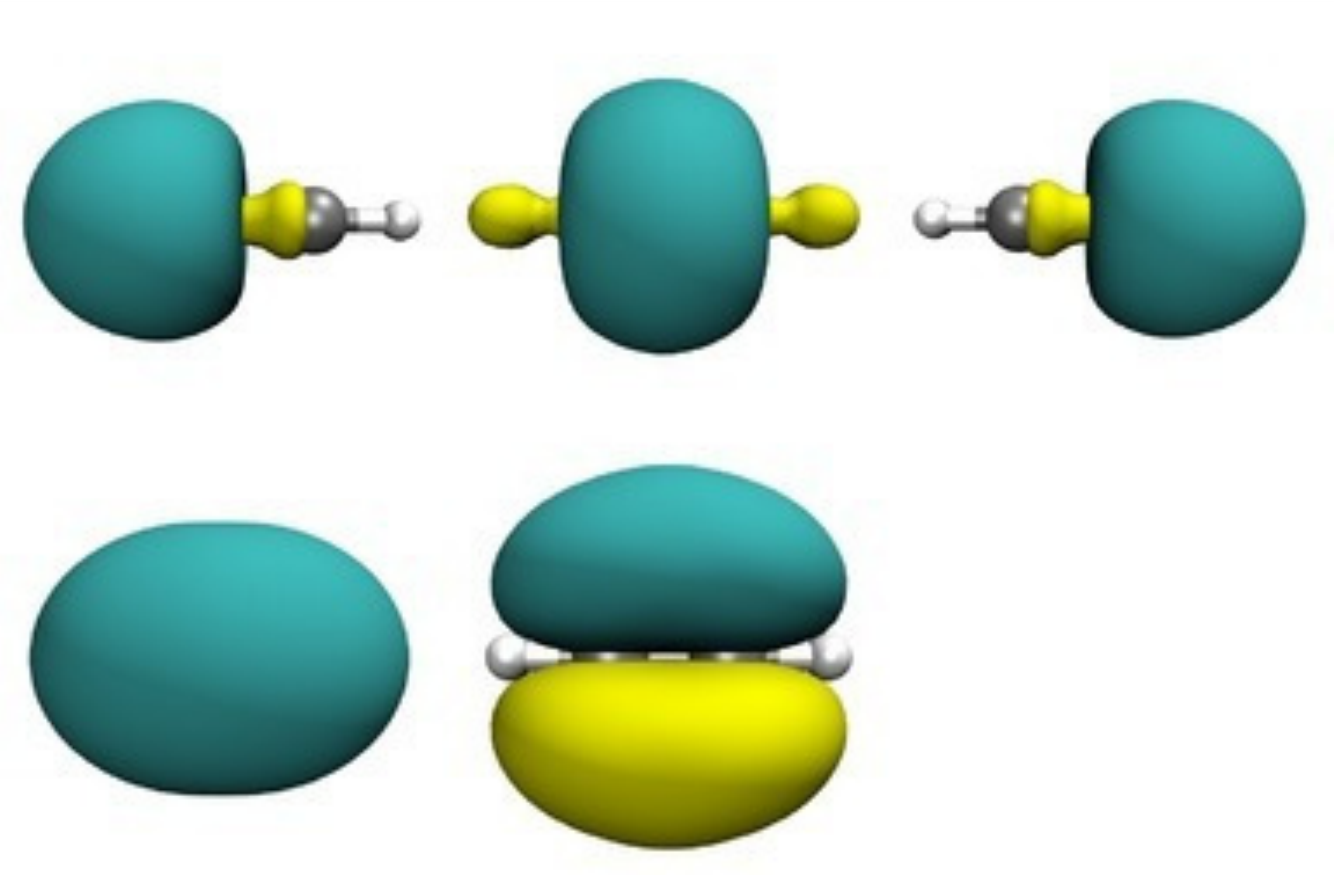}
{(a) $C_2H_2$}
\end{minipage}
\hspace{10pt}
\begin{minipage}[t]{0.22\textwidth}
\includegraphics[scale=0.28]{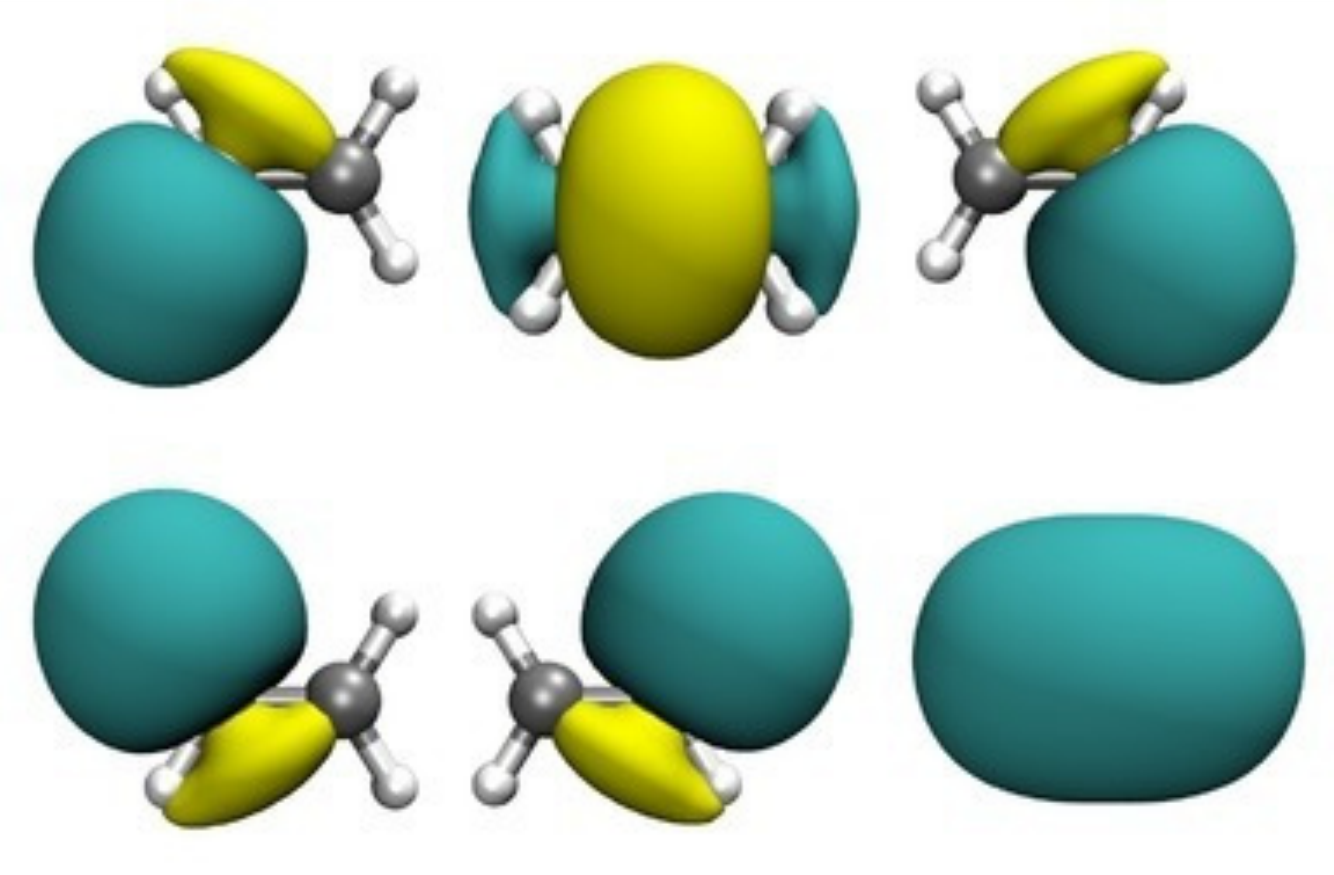}
{(b) $C_2H_4$}
\end{minipage}
\hspace{10pt}
\begin{minipage}[t]{0.28\textwidth}
\includegraphics[scale=0.29]{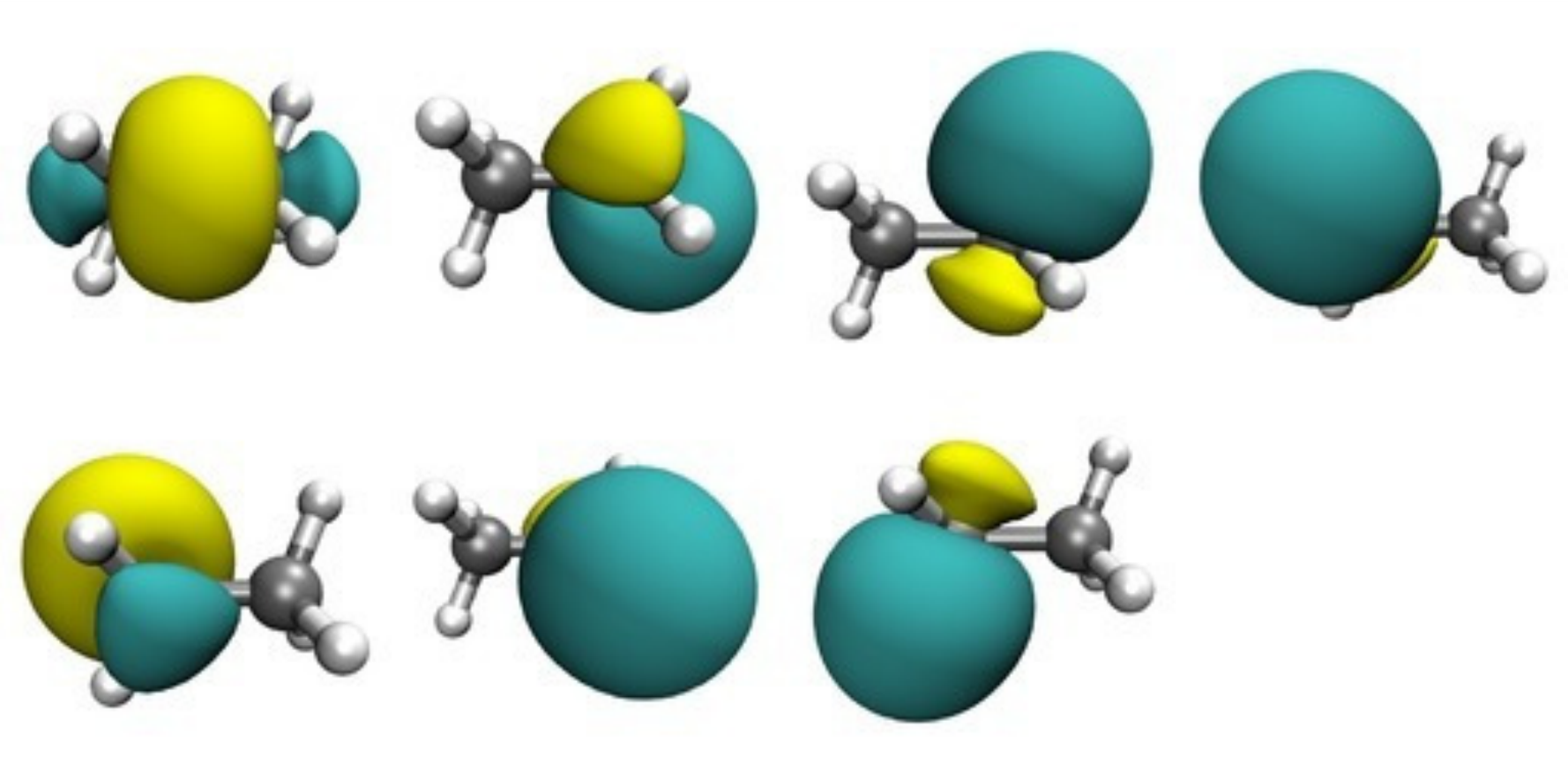}
{(c) $C_2H_6$}
\end{minipage}
\caption{\label{fig-WF-c2hn} The WFs of $C_2H_2$ (two $\pi_{CC}$
and one $\sigma_{CC}$ bonds), $C_2H_4$ (one $\pi_{CC}$ and one
$\sigma_{CC}$ bonds), and $C_2H_6$ (one $\sigma_{CC}$ bond).}
\end{figure}

The polarizability decomposition of acetylene ($C_2H_2$), ethylene
($C_2H_4$), and ethane ($C_2H_6$) was performed with
Wannier90~\cite{Mostofi2008} and a modified version of \QE
~\cite{Giannozzi2009} using the Perdew-Zunger
functional~\cite{Perdew1981} in a $35\times18\times18$ $a_0^3$
super-cell. The norm-conserving pseudo potential was used with energy
cutoffs of $40$ and $160$ Ry for the wavefunction and charge density,
respectively.  The maximally localized Wannier functions (MLWFs) of
these molecules are shown in Fig. \ref{fig-WF-c2hn}.  For $C_2H_2$ and
$C_2H_4$, a regular construction of WFs results in a mixture of $\pi$
and $\sigma$ bonds. Instead, $\sigma$ (from $p_x$) and $\pi$ (from
$p_y$ and $p_z$) orbitals were constructed separately, where $x$ is
along the direction of the $C$-$C$ bond.

\begin{figure}[ht]
\includegraphics[scale=0.9]{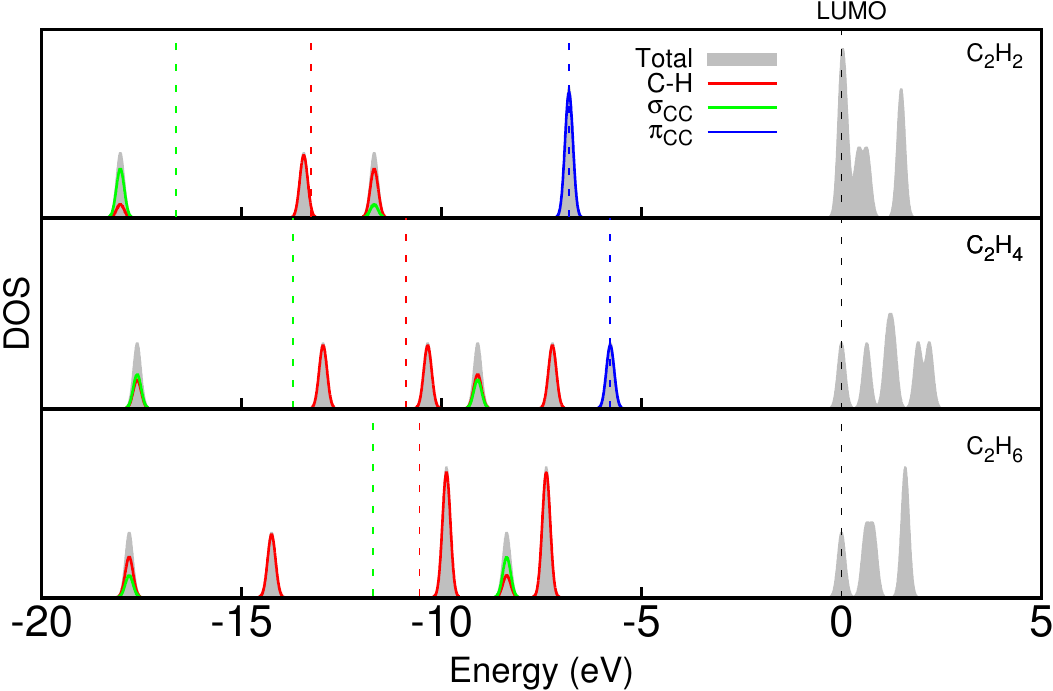}
\caption{\label{fig-PDOS-c2hn} Density of states projected onto WFs of
$C_2H_2$, $C_2H_4$, and $C_2H_6$. Colored dashed lines indicate the average
energy levels of Wannier orbitals.}
\end{figure}

The magnitude of the unscreened bond polarizability ($\bar{\alpha}^0$)
is inversely related to the transition energy ($\Delta E$) between the
occupied states and the onset of the unoccupied manifold, i.e., the
lowest unoccupied molecular orbital (LUMO). The relative energy of different
chemical bonds is indicated in the projected density of states (PDOS)
onto Wannier orbitals in Fig. \ref{fig-PDOS-c2hn}. Since WFs are not
eigenfunctions of the Kohn-Sham Hamiltonian, their PDOS can spread in
an energy range, making it hard to compare between different WFs. Here
we define the average energy of a WF as the on-site Hamiltonian
matrix element, $\langle W_n | H | W_n \rangle$, indicated by the
colored dashed lines in Fig.~\ref{fig-PDOS-c2hn}.  Within the same
molecule, in average $\Delta E(\pi_{CC}) < \Delta E(CH) < \Delta
E(\sigma_{CC})$, which implies that $\bar{\alpha}^0(\pi_{CC}) >
\bar{\alpha}^0(CH) > \bar{\alpha}^0(\sigma_{CC})$. It also explains
the increasing order of bond polarizabilities in
$C_2H_2$, $C_2H_4$, and $C_2H_6$ for the same kind of bonds.

\section{Locality of the response of Wannier functions}
The locality of $\Delta W_{{\bf
R}n,{\bf R}'n'}({\bf r})$ has two meanings as described in the main text. First $\Delta W_{{\bf
R}n,{\bf R}'n'} ({\bf r})$ decays exponentially with $|{\bf r}-{\bf
R'}|$ for a given pair of ${\bf R}$ and ${\bf R'}$. Second the
two-norm, $||\Delta W_{{\bf R}n,{\bf R}'n'}||$, decays exponentially
with $|{\bf R-R'}|$.

\begin{figure}[ht]
  \includegraphics[width=0.5\linewidth]{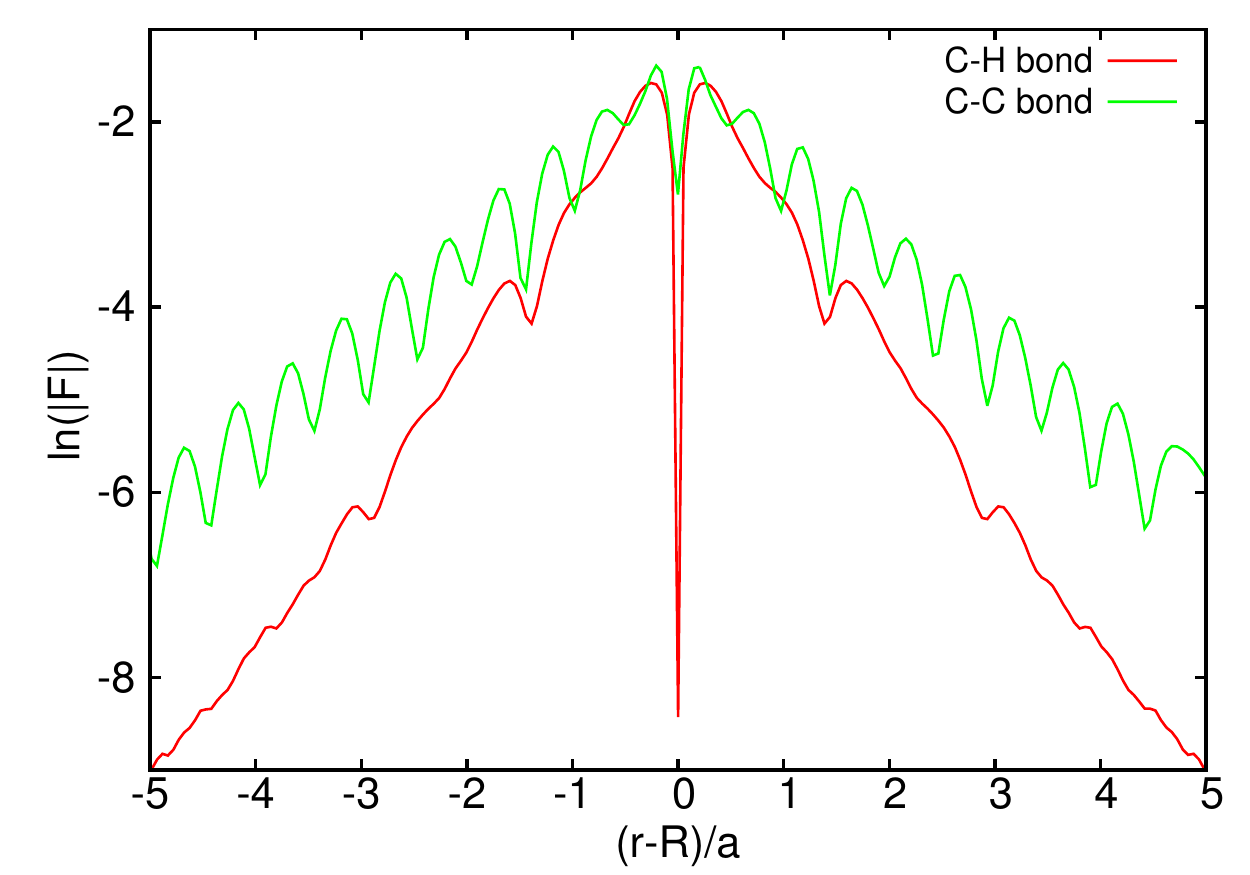}
  \caption{ \label{fig-Qvc} The exponential decay of $F({\bf r})$, which 
  results from the locality of both Wannier functions and the density matrix.
  }
\end{figure}

In fact, the first locality is closely related to that of the WFs and
the density matrix. As shown in Fig. \ref{fig-Qvc}, the quantity
$F({\bf r})=P_c\Delta V_sW_{Rn}({\bf r})$ exponentially decays in the real space,
and the decay length ($0.6a$ and $1.0a$ for C-H and C-C bonds,
respectively) is approximately twice as its corresponding WF ($0.3a$
and $0.5a$ for C-H and C-C bonds, respectively, not shown). This is
because the projector, $P_c({\bf r,r }') = I - P_v({\bf r,r }')$, has a similar
decay length as the WF, therefore broadens the spread of the WF by two
in $F({\bf r})$.  Comparing Fig.~\ref{fig-Qvc} and Fig. 1b in the main text,
we observe that the locality of $\Delta W_{{\bf R}n\bf{R'}n}({\bf r})$ is
similar to that of $F({\bf r})$, which means that the operator $[\tilde
\varepsilon - (H - \alpha P_v)I]^{-1}$ has very narrow spread in the
real space.

\begin{figure}[ht]
  \includegraphics[width=1\linewidth]{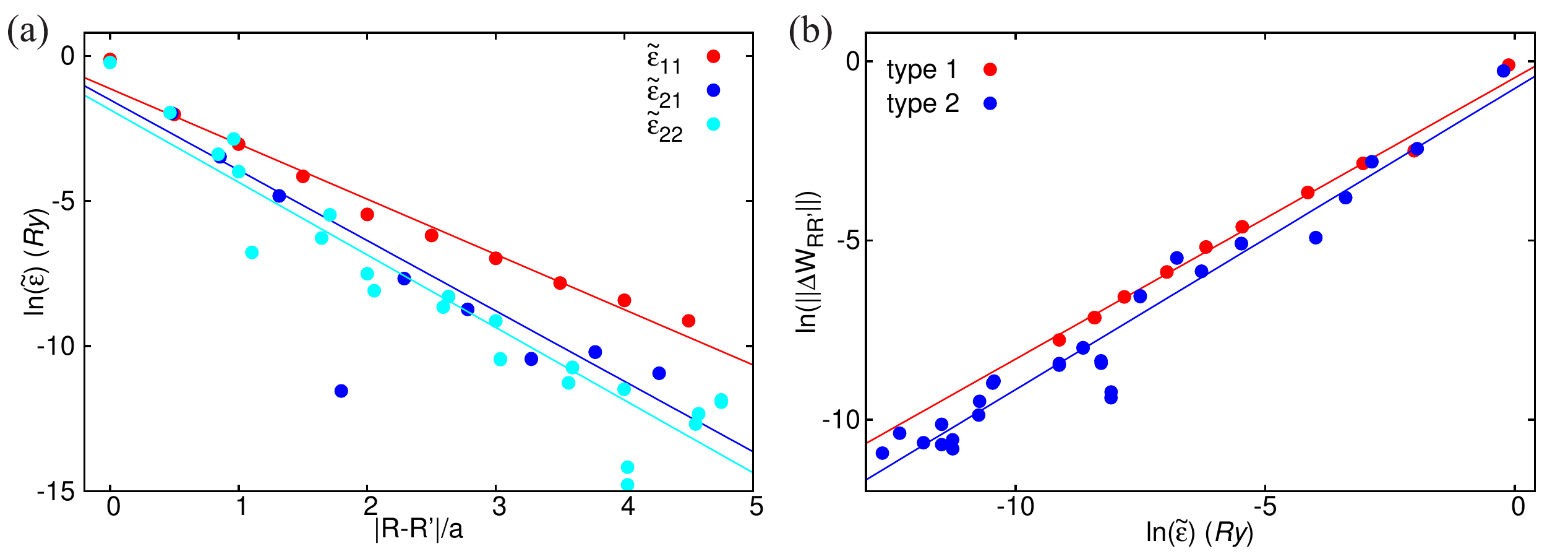}
  \caption{ \label{fig-EM} (a) The exponential decay of the
  Hamiltonian matrix in the Wannier basis. (b) The strong
  correlation between the magnitude of the local response and the
  hopping terms suggests that the rapid decay of the charge-flow is
  due to the decay of hopping terms between different WFs.}
\end{figure}

On the other hand, the decay behavior of $||\Delta W_{{\bf R}n,{\bf
R}'n'}||$ is dominated by the hopping term, $\tilde \varepsilon_{{\bf
R}n,{\bf R}'n'}$.  We show the exponential decay of $\tilde
\varepsilon_{{\bf R}n,{\bf R}'n'}$ in Fig.~\ref{fig-EM}a, and the
strong correlation between $\tilde \varepsilon_{{\bf R}n,{\bf R}'n'}$
and $||\Delta W_{{\bf R}n,{\bf R}'n'}||$ in Fig.~\ref{fig-EM}b.

\section{Scaling analysis of $\chi^0$}
In this section, we derive the numerical scaling to compute
$\chi^0$. We will only consider the static limit, as the computational
cost to calculate $\chi^0$ at an imaginary frequency follows the same
scaling. Here we focus on the computationally challenging case of
non-periodic extended systems. The computational model normally
contains a large number of electrons, $N$, in a super cell. For this
reason, we consider only the $\Gamma$-point in the Brillouin zone, and
drop ${\bf k}$ and ${\bf q}$ in the notation.  Under the plane-wave
implementation, suppose such a system has $N_v$ occupied states and
$N_c$ unoccupied states at a given kinetic energy cutoff. The number
of occupied Wannier orbital, $N_w$, is the same as $N_v$. The size of
the plane-wave grid is denoted by $N_G$. Here and in the following we
use capital $N$ for variables that scale with the system size, and use
lower case $n$ for variables independent of the system size.

The most straightforward way to compute $\chi^0$ is to use the
Adler-Wiser formula~\cite{ADLE62,WISE63}:
 \begin{equation}
    \chi^0_{\bf G,G'} = \frac{1}{\Omega}\sum_{ij} 
                      (f_i-f_j)
                      \frac{
                        \langle i  | e^{-i{\bf G\cdot r}}  | j \rangle
                        \langle j | e^{i{\bf G'\cdot r'}}  | i \rangle }
                      {\varepsilon_{i}-\varepsilon_{j}+i\eta}.
    \label{AW}
 \end{equation}
For semiconductors and insulators, Eq.~\ref{AW} requires $N_v \times
N_c$ Fast Fourier Transforms to calculate all pairs of co-density
$\psi_i^*({\bf r})\psi_j({\bf r})$ in the $G$ space, whose cost is proportional to
$N_v \times N_c \times N_GlnN_G$.  Evaluating the full matrix of
$\chi^0$ requires four loops for ${\bf G,G'},i$, and $j$, so the
leading order of the scaling is $O(N^4)$. Besides, the direct approach
needs to store $N_G^2$ elements. Both the computational and storage
expenses become major bottlenecks for large systems.
 
Alternatively, as illustrated in the main text, we could represent
$\chi^0$ in a local basis set, $\{| \xi^n_i \rangle\}$, made of
approximate eigenmodes of local response.  $| \xi^n_i \rangle$ refers
to the $i$-th approximate eigenvector of the local response of the
$n$-th WF, $\chi^0_{nn}$. If we need $n_m$ eigenmodes for each WF, the
cost of the storage is proportional to $n_m\times N_w \times N_G$.
Since $n_m$ is independent of the system size, the cost to store
$\chi^0$ grows as $O(N^2)$, but with a much smaller prefactor as
compared to the direct evaluation of the Adler-Wiser formula.

Now we analyze the computational cost of the local basis set
approach. In general $\{|\xi_i^{n}\rangle\}$ are obtained by
iteratively solving the generalized Sterheimer equation and update
$\Delta V_s$ for several times (typical in the order of ten), until
the new $\Delta V_s$ is an eigenstate of $\chi^0$. The matrix form of
this equation reads:
  \begin{equation}
      \left ( \begin{array} {ccc}
      \tilde\varepsilon_{0,0}-H & \dots &  \tilde\varepsilon_{0,N_w}\\
      \vdots& \ddots & \vdots  \\
      \tilde\varepsilon_{N_w,0}&\dots&\tilde\varepsilon_{N_w,N_w}-H \\
     \end{array} \right ) 
     \left ( \begin{array} {c}
      \Delta W_{0} \\
      \vdots        \\
      \Delta W_{n_w} \\
     \end{array} \right ) = 
     \left ( \begin{array} {c}
      P_c\Delta V_s W_{0} \\
      \vdots        \\
      P_c\Delta V_s W_{n_w} \\
     \end{array} \right ) .
     \label{matrix-gse}
  \end{equation}
We construct the local basis set by removing the off-diagonal elements of $\tilde \epsilon$, and solve
the eigenvectors of 
\begin{equation}
(\tilde\varepsilon_{i,i}-H)  \Delta W_{i} =P_c\Delta V_s W_{i}. \label{eq-decouple}
\end{equation}
To apply projector $P_c=I-\sum_i |W_i\rangle\langle W_i|$ in principle
requires $N_v\times N_G$ operations, however, we can benefit from the
exponential locality of $W_{i}$, and truncate the summation
of the Wannier centers to $n_n$ neighbors within a properly chosen
cutoff distance of the local perturbation without losing accuracy.
The computational cost to calculate the right hand side of
Eq.~\ref{eq-decouple} therefore scales as $n_n\times N_G$, and the
iterative solution of $\Delta W_{i}$ scales as $N_G \ln N_G$. The cost
to construct the full local basis set will acquire a prefactor of $n_m
\times N_w$, which gives an overall $O(N^2)$ scaling.

To compute $X_{ij}^{nn'} =  \langle \xi_i^{n} | \chi^0 | \xi_j^{n'} \rangle$ used in the dielectric band structure interpolation, 
one has to solve $N_w$ coupled equations in Eq.~\ref{matrix-gse}.
Again, thanks to the exponential localization of $\chi^0$, only $n_n$ coupled equations needs to be solved for each $|\xi_j^{n'} \rangle$.
Therefore the leading cost to construct $X$ is $n_n$ times of the cost of building the basis set, which again grows as $O(N^2)$.
Compared to the $O(N^4)$ scaling of the standard Adler-Wiser formula, the local basis set approach has a tremendous saving in both computer time and memory.

\bibliographystyle{h-physrev}